\documentclass[12pt]{article}
\usepackage{amsmath}
\usepackage{amssymb}
\numberwithin{equation}{section}
\usepackage[left=0.98in,right=0.98in,bottom=1.5in]{geometry}
\usepackage{setspace}
\usepackage [linktocpage]{hyperref}
\begin{document}
\title{\bf Dual Localized Objects From \\ M-Branes Over $AdS_4 \times S^7/Z_k$ \ \\ \ }
\author{{\bf M. Naghdi \footnote{E-Mail: m.naghdi@mail.ilam.ac.ir} } \\
\textit{Department of Physics, Faculty of Basic Sciences}, \\
\textit{University of Ilam, Ilam, West of Iran.}}
\date{\today}
 \setlength{\topmargin}{0.1in}
 \setlength{\textheight}{9.2in}
  \maketitle
  \vspace{-0.0in}
    \thispagestyle{empty}
    \begin{center}
\textbf{Abstract}
 \end{center}
We consider a few ansatzs for the four and seven forms of 11-dimensional supergravity over $AdS_4 \times S^7/Z_k$ while try to keep the geometry unchanged. From the 4-form equations, we arrive at some massless scalars and pseudoscalars in the bulk of Euclidean $AdS_4$ that match with some boundary $\Delta_+=3$ operators. Indeed, the main objects are instantons and domain walls as the fully and partially localized objects in the external space, respectively. The latter comes from an (anti)M5-brane wrapping partly around three internal directions similar to the fuzzy $S^3/Z_k$ solutions. Except for the first solutions of adding some anti-M2-branes$\mid$(M2-branes) to the original M2-branes, the objects backreact on the geometry although small$\mid$(break all supersymmetries and destabilize the vacua). The dual 3-dimensional field theory solutions are got by the skew-whiffing $\textbf{8}_s \rightarrow \textbf{8}_v$ and $\textbf{8}_s \rightarrow \textbf{8}_c$ for the scalars and pseudoscalars respectively, while the gauge fields are used mainly for the $k=1,2$ cases where the R-symmetry and supersymmetry are enhanced as $SU(4)\times U(1)\rightarrow SO(8)$ and $\mathcal{N}=6 \rightarrow \mathcal{N}=8$ respectively, and also for pseudoscalars. Further, for the pseudoscalars we propose a special boundary deformation, with a fermion field, that is equivalent to a multi-trace deformation already studied for the bulk $m^2=-2$ conformally coupled pseudoscalar.

\newpage
\setlength{\topmargin}{-0.7in}
\pagenumbering{arabic} 
\setcounter{page}{2} 

\section{Introduction}
In general, $\mathcal{N}=8$ gauged supergravity in 4 dimensions is a consistent Kaluza-Klein reduction from 11-dimensional (11D) $\mathcal{N}=1$ supergravity on $S^7$ \cite{Duff84NPW} so that any solution to the low-dimensional equations of motion (EOMs) could be uplifted to the full high-dimensional ones. It is also discussed in \cite{Gauntlett} that, in this way, a finite numbers of states including massless graviton, gauge fields and matter fields (not necessarily massless) are always kept, and the simplest way to achieve a consistent truncation is to keep all singlet fields under a symmetry group and setting other nonsinglet ones to zero. By the way, here we consider the standard $N$ M2-brane theory, by Aharony-Bergman-Jafferis-Maldacena (ABJM), on $Z_k$ orbifold of $C^4$ that has the near horizon geometry of $AdS_4 \times S^7/Z_k$ with $N$ units of the 4-form flux in the bulk of $AdS_4$, and the legality limit of $N\gg k^5$ (the weakly curved space). In the case, one often considers $S^7/Z_k$ as a $S^1/Z_k$ Hopf-fibration on $CP^3$, and when $k$ increases the radius of the M-theory circle decreases and a better description is for $N$ D2-branes of type IIA supergravity over $AdS_4 \times CP^3$ with associated fluxes. The dual field theory is a 3d $\mathcal{N}=6$ conformal Chern-Simons-matter $SU(N)_k \times SU(N)_{-k}$ gauge theory with a $SU(4)$ R-symmetry and the matter fields transforming in bifundamental representations of the gauge group, where $k$ is the Chern level \cite{ABJM}. For the cases $k=1,2$ and $N=2$ the R-symmetry and supersymmetry are enhanced to $SO(8)$ and $\mathcal{N}=8$ because of monopole operators respectively, which the latter is indeed the Bagger-Lambert-Gustavsson (BLG) model--Look, for instance, at \cite{1203.3546} as a general reference.

We have already studied some exacts solutions in M2/D2-branes theory as instances of nonperturbative effects in the gauge and gravity theories. Depending on how many spatial dimensions the objects have in the bulk, we call them instantons, monopoles or vortexes, strings, and so on. By instantons we mean fully localized objects in the external Euclidean space ($EAdS_4$ here) with finite actions, which enter the tunneling processes among various vacua and have many important roles from particle physics up to early universe cosmological theories. In \cite{I.N} we found an instanton agreeing to a bulk pseudoscalar with $m^2=-2$ in anti-membrane theory; while in \cite{N}, and \cite{I}, an $U(1)$ (monopole-)instanton was surveyed. Next, in \cite{Me3} and \cite{Me4}, we included some (anti)D2- and (anti)D4-branes in the type IIA limit of the ABJM theory and fixed their boundary dual according to the AdS/CFT, specially state-operator, correspondence rules \cite{KlebanovWitten}. We follow mainly the studies in the last two references in searching to find new localized objects in the M2-branes theory with new proposals.

Here we add some new ansatzs for the 4- and 7-form field to the 11D supergravity (the so-called Freund-Rubin) ones of ABJM over $AdS_4 \times S^7/Z_k$, without deforming the background geometry, \footnote{It is notable that we consider the branes as solutions to the supergravity EOMs with appropriate source terms, where these (always delta function) terms come from the coupling of the brane actions to the main supergravity action. In other words, one may write the full gravitational action as $S=S_{SUGRA}+S_{DBI}+S_{CS}$ in which any fluctuation for the existing or embedded branes can be realized through the latter two terms. Our procedure is similar in that we are indeed working on the main $S_{SUGRA}$ action and by adding some new terms to the background ones, the dynamics of the existing branes or that of the new included ones are surveyed. As long as the new ansatzs for the form fields, combined with the original ones, satisfy the equations and identities of the main theory, the procedure is right and one may obtain consistent solutions. One should also note that although we try to keep the main geometry unchanged, because of the gravity coupled to the other involved fields, there are always some backreactions on the background geometry.} and find some fully or partially localized massless scalars and pseudoscalars in the bulk. The solutions may be interpreted as instantons, strings and domain walls, based on the spatial dimensions they have in the external space. In fact, our second included electric (anti)M2-brane wraps in just one internal direction while the third one wraps partly around three internal directions resulting in a string of monopole-instantons along the horizon. To find the corresponding boundary solutions, we should employ both gauge and matter (scalar and fermion) fields. In the procedure, we swap the representations for the gravitino of the original M2-branes theory, which is in turn performed by knowing that there are some triality transformations among three representations ($\textbf{8}_s, \textbf{8}_c, \textbf{8}_v$) of the $SO(8)$ internal isometry group. Actually, to find the dual scalars and pseudoscalars, we exchange $\textbf{8}_s \rightarrow \textbf{8}_v$ and $\textbf{8}_s \rightarrow \textbf{8}_c$ respectively, while the favorite singlet gauge field is in the original representation. For duals to the bulk scalars, we use just one boundary scalar and a dimension-3 operator with the same structure as the sextic scalar potential of the original boundary field theory, while for pseudoscalars we deforms the corresponding action with a suitable $SU(4)_R \times U(1)_b$-singlet conformal-breaking operator made of just one fermion with the $U(1)$ gauge fields turned on. Meanwhile, the main ansatzs do not preserve conformal symmetry and any supersymmetry and also destabilize the vacua as we will describe briefly.

The structure of the remaining parts of this note is as follows. In Section 2, we deal with the gravity side of the study. In subsection 2.1, we write the background materials including geometry, fields and equations of motion in our Euclidean notation. In the next three subsections, we present our ansatzs, solutions and some related issues while in the last 2.5 subsection, we discuss on symmetries and supersymmetries that the solutions have. In Section 3, we go on the field theory side of the study. In three subsections there, we present plain dual boundary scalars, gauges fields and pseudoscalars to match with the bulk solutions with needed arguments. In Section 4, we discuss further on the solutions with relating them to the previous related studies and comment on supersymmetry breaking and instability because of the solutions.

\section{On Super-gravity Side}
\subsection{The Background}
We use the metric of 11D supergravity over $AdS_4 \times S^7/Z_k$ as
\begin{equation}\label{eq01}
ds^2_{M}=\frac{R^2}{4} ds^2_{EAdS_4}+R^2 ds^2_{S^7/Z_k},
\end{equation}
where $E$ in $EAdS_4$ is for Euclidean, $R=R_7=2R_{AdS}$ is the curvature radius of the 11D target-space,
\begin{equation}\label{eq01a}
\ \ \ \ \ \ \ ds^2_{EAdS_4}=\frac{1}{u^2} \big(du^2+ dx_i dx_i \big), \quad i=1,2,3, 
\end{equation}
\begin{equation} \label{eq01b}
ds_{S^7/Z_k}^2 =ds_{CP^3}^2+\frac{1}{k^2}(d\varphi+k\omega)^2,
\end{equation}
and
\begin{equation}\label{eq01cc}
  \begin{split}
   ds_{CP^3}^2 & = d\xi^2 + cos^2\xi sin^2\xi \left(d\psi + \frac{1}{2} cos\theta_1 d\varphi_1 - \frac{1}{2} cos\theta_2 d\varphi_2 \right)^2 \\
   & + \frac{1}{4} cos^2\xi \big(d\theta_1^2 + sin^2\theta_1 d\varphi_1^2 \big) + \frac{1}{4} sin^2\xi \big(d\theta_2^2 + sin^2\theta_2 d\varphi_2^2 \big)
  \end{split}
\end{equation}
is for $CP^3$ metric with the angles $0 \leq \xi \leq \pi/2, \, 0 \leq \theta_s \leq \pi, \, 0 \leq \varphi_s, \varphi, \psi \leq 2\pi, \, s=1, 2$, and
\begin{equation} \label{eq01d}
 \omega=\frac{1}{2} \big((cos^2\xi-sin^2\xi) d\psi+cos^2\xi cos\theta_1 d\varphi_1+ sin^2\xi cos\theta_2 d\varphi_2 \big)
\end{equation}
is a topologically nontrivial 1-form related to the K$\ddot{a}$hler form $J (=d\omega)$ on $CP^3$. Note that we are indeed using the case when one considers $S^7/Z_k$ as a $S^1/Z_k$ Hopf-fibration (with $\acute{\varphi}=\varphi/k$ as the $S^1$ fiber coordinate) on $CP^3$. \footnote{We remind that $k$, as the Chern level in the filed theory side, is the orbifold parameter labeling various $Z_k$ quotients. In fact, with four complex coordinates $y_A$ $(A=1,2,3,4)$ for the transverse space to M2-branes, the $Z_k$ quotient acts as $y_A = e^{i\frac{2\pi}{k}} y_A$. Specially, for $k=1,2$, the theory describes M2-branes in flat $R^8$ space and probing an $R^8/Z_k$ singularity respectively and that with increasing $k$, just the $CP^3$ space remains. Here we use a general $k$ in our analyses and computations of solutions, charges and actions.} Besides, for the future purposes, we define the frame 1-forms (vielbeins)
\begin{equation}\label{eq01e}
\begin{split}
& e^1=d\xi, \quad e^2=\frac{1}{2} cos\xi d\theta_1, \quad e^3=\frac{1}{2} cos\xi sin\theta_1 d\varphi_1, \quad e^4=\frac{1}{2} sin\xi d\theta_2, \quad e^5=\frac{1}{2} sin\xi sin\theta_2 d\varphi_2, \\
& \ \ \ \ \ \ \ \ \ e^6= cos\xi sin\xi \big(d\psi+\frac{1}{2}cos^2\xi cos\theta_1 d\varphi_1-\frac{1}{2} cos\theta_2 d\varphi_2 \big),\quad e^7=\frac{1}{k}(d\varphi+k\omega)\equiv e_7
\end{split}
\end{equation}
for the internal $S^7/Z_k$ space. Meanwhile, for the corresponding 4-form field strength we use
\begin{equation}\label{eq02}
  G_4^{(0)}=d\mathcal{A}_3^{(0)} = + i \frac{3}{8} R^3 \mathcal{E}_4=+ i N \mathcal{E}_4,
\end{equation}
where $\mathcal{E}_4$ is the unit-volume form on $EAdS_4$, and to have $N$ units of the 4-form flux on the quotient space, $\acute{N}=k N$, which  is in turn the number of the flux quanta on $S^7$, and note also that the validity of the 11D supergravity approximation is when $N \gg k^5$. One should also note that (\ref{eq02}) is actually the Euclideanized version of the ABJM background \cite{ABJM} when we go form the \emph{almost positive} Minkowskian signature to the \emph{fully positive} Euclidean signature after the Wick rotation $t\rightarrow + it_E$, and that with just a minus sign therein we have the \emph{skew-whiffed} background of the Minkowskian signature.

On the other hand, the Euclideanized action of the 11D supergravity, with our formalism as $S^{Minkoski} \rightarrow i S^{E}$, reads
\begin{equation}\label{eq03}
  S_{M}^E = -\frac{1}{2 \kappa_{11}^2} \biggl\lbrack \int d^{11}x \, \sqrt{g} \, \mathcal{R} + \frac{1}{2} \int \biggl(G_4 \wedge \ast G_4 - \frac{i}{3} \mathcal{A}_3 \wedge G_4 \wedge G_4 \biggr)\biggr\rbrack,
\end{equation}
with $G_4=d\mathcal{A}_3$, the Hodge-dual $*$ with respect to the full 11D metric,  and $\kappa_{11}^2=8\pi  \mathcal{G}_{11}=\frac{1}{4\pi}(2\pi l_p)^9$, where $ \mathcal{G}_{11}$, $\kappa_{11}$ and $l_p$ are the 11D Newton's constant, gravitational constant and Plank length, respectively. From the action, the EOM for $\mathcal{A}_3$ reads
\begin{equation}\label{eq04a}
  d\ast G_4 - \frac{i}{2} G_4 \wedge G_4=0,
\end{equation}
and the Bianchi identity is $dG_4=0$. Moreover, from the metric variation, we get
\begin{equation}\label{eq04b}
    \mathcal{R}_{MN}-\frac{1}{2} g_{MN} \mathcal{R}=8\pi \mathcal{G}_{11} T_{MN}^{G_4},
\end{equation}
in which
\begin{equation}\label{eq04b1}
 T_{MN}^{G_4}=\frac{1}{6} \biggl[G_{MPQR} G_N^{PQR}-\frac{1}{8} g_{MN} G_{PQRS} G^{PQRS} \biggl],
\end{equation}
where we use the capital indices $M, N,...$ for the 11D space-time directions.

Further, to evaluate the electric (page) and magnetic (topological) charges associated with M2-branes and M5-branes respectively, we use
\begin{equation}\label{eq050}
 Q_e=\frac{1}{\sqrt2 \kappa_{11}^2} \int \big(\ast G_4 +\frac{1}{2} \mathcal{A}_3 \wedge G_4 \big)= \frac{1}{\sqrt2 \kappa_{11}^2} \int G_7, \qquad Q_m=\frac{1}{\sqrt2 \kappa_{11}^2} \int G_4.
\end{equation}

\subsection{The First Ansatzs and Solutions}
The first sets of ansatzs, already used for the 4-form of type IIA supergravity over $AdS_4 \times CP^3$ in \cite{Me4} as well, here are
\begin{equation}\label{eq05}
      G_4^{(1)}=i d({f_1}^{-1}) \wedge dx \wedge dy \wedge dz,
\end{equation}
\begin{equation}\label{eq06}
     G_4^{(2)}= i df_2 \wedge \mathcal{A}_3^{(0)} + i f_2 G_4^{(0)},
\end{equation}
with $i$ factor because of being in Euclidean space, and $f_1,f_2,...$ are some (pseudo)scalar functions in the bulk of $EAdS_4$--They are real here in that come from the external components of $\mathcal{A}_{MNP}$. The Bianchi identity is valid for the ansatzs trivially; and from the first term of Eq. (\ref{eq04a}), $d(\ast_4 G_4^{({s})})=0$, we simply obtain
\begin{equation}\label{eq05b}
      \frac{d^2f_1(u)}{du^2} -\frac{2}{f_1(u)} \big(\frac{df_1(u)}{du} \big)^2 + \frac{4}{u}\frac{df_1(u)}{du}=0 \Rightarrow f_1(u)=\frac{3u^3}{c_1-3c_2u^3},
\end{equation}
\begin{equation}\label{eq06b}
   d\ast_4 \big(df_2 \wedge \mathcal{A}_3^{(0}\big)+\frac{6}{R} df=0 \Rightarrow  \frac{d^2f_2(u)}{du^2} -\frac{2}{u}\frac{df_2(u)}{du}=0 \Rightarrow f_2(u)=c_3+c_4 {u^3},
\end{equation}
where $c_1, c_2,...$ are some bulk constants with physical meaning and we have used the distributive property of the Hodge-star because of the diagonal metric, and also the conventions
\begin{equation}\label{eq07}
\begin{split}
  & \sqrt{g}= \sqrt{g_4} \sqrt{g_7}=\big(\frac{R}{2}\big)^4 . R^7=\frac{R^{11}}{16}, \qquad \ast_4 \mathcal{E}_4= \frac{16}{R^4}, \qquad \ast_7 \mathcal{E}_7= \frac{1}{R^7}, \\
   &\ \ \ \ \ \ \ \ \ \ \mathcal{E}_4=\frac{1}{u^4} dx \wedge dy \wedge dz \wedge du, \qquad \mathcal{E}_7=\frac{1}{8.3!} J^3 \wedge e_7,
   \end{split}
\end{equation}
with $e_7$ as the seventh vielbein that we defined it in (\ref{eq01e}).

Now, with the fact that in the ABJM M2-branes theory \cite{ABJM}
 \begin{equation}\label{eq08}
R/l_p = (2^5 \pi^2 \acute{N})^{1/6} \Rightarrow \kappa_{11}^2=\frac{16 \pi^5}{3} \sqrt{\frac{R^9}{3k^3}},
\end{equation}
the M-branes electric charges, in the unit internal $S^7/Z_k$ volume, based one the solutions (\ref{eq05b}) and (\ref{eq06b}) from (\ref{eq050}) read
\begin{equation}\label{eq08a}
 Q_e^1=\frac{3 c_1}{\pi^5} \sqrt{\frac{3 k^3}{2 R^9}}, \qquad Q_e^2= \frac{9 c_3}{8 \pi^5} \sqrt{\frac{3 k^3}{2 R^{11}}},
\end{equation}
respectively. Similarly, the corresponding magnetic charges, in the unit external $EAdS_4$ volume, can be easily evaluated which are equal to the same electric charges in the unit-volume of the internal 7D space.

Next, by plugging $G_4=G_4^{(0)}+G_4^{({s})}$ into the action (\ref{eq03}) and noting that the Chern-Simons term vanishes in the case, for the corrections to the background action, in the unit 11D volume, we obtain
\begin{equation}\label{eq09}
  S_{M.1,2}^{E} = \frac{3}{8 \pi^5} \sqrt{\frac{3 k^3}{R^{15}}}\big(64\tilde{c_1}^2 \pm 48\tilde{c_1}+9\big),
\end{equation}
with $c_1=\tilde{c_1} R^3$ based on the solution (\ref{eq05b}) and with $c_3=\frac{8}{3} \tilde{c_1}$ based on the solution (\ref{eq06b}). Now, one should note that with the upper ($+$) sign and $\tilde{c_1}=\frac{3}{8}$, we are indeed adding some M2-branes with the same properties as the original $N$ M2-branes, while for the lower ($-$) sign and the same $\tilde{c_1}$, corresponding to the skew-whiffing, or anti-M2-branes added on the original M2-branes, the action value vanishes (something like pair annihilations!).

Further, for the backreaction issue, we note that the ansatzs (of the Freund-Rubin type) have the same structure as the background one (\ref{eq02}) and so with
\begin{equation}\label{eq10a}
G_4^{(tot,1)}= + i \frac{3}{8} R^3 \mathcal{E}_4 + i c_1 \mathcal{E}_4, \qquad
G_4^{(tot,2)}= + i \frac{3}{8} R^3 \mathcal{E}_4 + i \frac{3}{8} R^3 c_3 \mathcal{E}_4,
\end{equation}
based on the corresponding solutions, the condition to have a zero correction to the background energy-momentum tensor comes from vanishing the second sentence of (\ref{eq04b1}), which is in turn proportional with $G_4^{(tot.s)} \wedge \ast G_4^{(tot.s)}=0$, with the result
\begin{equation}\label{eq10}
  \left\{ \begin{split}
   & c_1=-\frac{3}{4} R^3, \\
   & c_3=-2,
  \end{split} \right. \Rightarrow G_4^{(s)}= - i \frac{3}{4} R^3 \mathcal{E}_4,
\end{equation}
that means, to avoid  backreactions, we must add the special anti-M2-branes to the original M2-branes and conversely. \footnote{Note is required that one could also think of the solutions as some fluctuations on top of the original background one or its skew-whiffed version and then, to elude the backreactions one have to consider the condition (\ref{eq10}) just reached on the solutions.}

\subsection{The Second Ansatz and Solution}
The next new ansatz we employ reads
\begin{equation}\label{eq11}
      G_4^{(3)}= i (\ast_4 df_3) \wedge d\varphi \Rightarrow \ast G_4^{(3)} \equiv G_7^{(3)} = - i df_3 \wedge (\ast_7 d\varphi),
\end{equation}
with a note that $G_4^{(3)}$ may be associated to an electric (anti)M2-brane wrapping one internal and two external directions, while the $G_7^{(3)}$ may be associated to its dual (anti)M5-brane wrapping the other six internal directions and so, resulting in a pseudoscalar.
From the EOM (\ref{eq04a}), we obtain
\begin{equation}\label{eq12}
   \ast_4 d(\ast_4 df_3)=d^{\dagger} df_3=\frac{1}{\sqrt{g_4}} \partial_{\acute{\mu}} \big(\sqrt{g_4}\ g^{\acute{\mu}\acute{\nu}} \partial_{\acute{\nu}} f_3 \big)=0,
\end{equation}
where $\acute{\mu}, \acute{\nu},...$ are for the bulk indices, with the solution (look at \cite{Park} for a derivation)
\begin{equation}\label{eq12b}
      f_3(u,\vec{u};0,\vec{u}_0)=c_5 + \frac{c_6 u^3}{[u^2+(\vec{u}-\vec{u}_0)^2]^3},
\end{equation}
which is a propagator from the boundary to the bulk.

To evaluate the associated charges, we use
\begin{equation}\label{eq13}
\int df_3=-\frac{c_6}{9 \epsilon^3}, \qquad \int \ast_4df_3=\frac{\pi c_6}{32} \frac{R^2}{\epsilon^3},
\end{equation}
evaluated based on the solution (\ref{eq12b}), where to have finite integrals we set $x_{\acute{\mu}}=\epsilon \neq 0$. Therefore, the magnetic charge from (\ref{eq050}) with (\ref{eq11}) and (\ref{eq13}) reads 
\begin{equation}\label{eq150}
 Q_m^3 = \frac{3}{256 \pi^3} \frac{c_6}{\epsilon^3} \sqrt \frac{3 k^3}{2 R^{5}},
\end{equation}
for the finite part. But for the associated electric charge, we need to evaluate $\ast_7 d\varphi$. So, for simplicity, by writing the 1-form $\omega$ in (\ref{eq01d}) in the orthogonal frame, making use of $e_7$ in (\ref{eq01e}), and noting that
\begin{equation}\label{eq14}
\ast_7 e^a=\frac{1}{6!}\ \varepsilon^a_{\ bc...7}\ e^b \wedge e^c \wedge ... \wedge e^7, \quad \varepsilon_{1234567}=+1, \quad e^a=e^a_m dx^m,
\end{equation}
where $a, b,...$ and $m, n,...$ stand for the tangent and space-time internal 7D indices respectively, and $\varepsilon_{mn...7}$ is the antisymmetric Levi-Civita tensor, we arrive at
\begin{equation}\label{eq15}
 Q_e^3 = \frac{c_6}{384 \pi^6} \frac{4-k}{\epsilon^3} \sqrt \frac{k^3}{R^{13}}
\end{equation}
as the finite part of the charge, with $k=1$ for the $\acute{\varphi}$ integration.

The correction to the original action can also be calculated with the ansatz (\ref{eq11}) with (\ref{eq03}), by evaluating the external and internal components separately. For the external component, we get
\begin{equation}\label{eq16b}
\int_{EAdS_4} df_3 \wedge \ast_4 df_3=\frac{21\pi^2}{1024} \frac{c_6^2}{\epsilon^6} R^2,
\end{equation}
which is the finite contribution with $\epsilon \neq 0$ as the lower limit of the $u$ integration and considering the boundary as a 3-shpere at infinity, $S^3_\infty$. Further, for the external component, with
\begin{equation}\label{eq16c}
d\varphi \wedge \ast_7 d\varphi= \bigg(1+\frac{1}{4sin^2\xi}+\big(cos\xi-\frac{1}{2}cot\theta_1\big)^2+\big(sin\xi cot\theta_2+\frac{1}{2}cot\xi cot\theta_2\big)^2 \bigg)\ \mathcal{E}_7,
\end{equation}
and $\int J^3 \wedge e_7=(2\pi)^4/k$, the finite contribution that we get reads
\begin{equation}\label{eq16d}
  \int_{S^7/Z_k} d\varphi \wedge \ast_7 d\varphi=\frac{3 k \pi^4}{4} R^5.
\end{equation}
Therefore, the whole correction to the action becomes
\begin{equation}\label{eq17}
  S_{M.3}^{E} \cong - 0.002 \frac{\pi c_6^2}{\epsilon^6} \sqrt{k^5 R^7},
\end{equation}
that is a very small contribution for finite $k$ and $R$.

Besides, to handle the backreaction on the geometry because of the solution, one should note that with
\begin{equation}\label{eq18}
 \begin{split}
     G_4^{(tot.3)} & = i N \mathcal{E}_4 + i (\ast_4 df_3) \wedge d\varphi = \frac{1}{4!} G_{MNPQ}^{(tot.3)}\ dX^{MNPQ}, \\
      & \Rightarrow G_{MNPQ}^{(tot.3)} = i\frac{3 R^3}{8} \varepsilon_{\acute{\mu}\acute{\nu}\acute{\rho}\acute{\sigma}}- i \frac{R^2}{4}(\varepsilon_{\acute{\mu}\acute{\nu}\acute{\rho}}^{\ \ \ \ \acute{\lambda}}\ \partial_{\acute{\lambda}}f_3), 
 \end{split}
\end{equation}
the excess to the main contribution can be evaluated from (\ref{eq04b1}), which does not vanish, except in leading order of the solution expansion, for both internal and external components because of $d(f_3 \ast_4 df_3)\neq 0$. \footnote{It is remarkable that we may be able to ignore the backreactions because of the second (and the third) solution according to already known facts in similar situations. The first is that as long as we are interested in studying the solution behaviours near the boundary and correlation functions of the dual boundary operator for the active (pseudo)scalar, and the gravity-scalar equations decouple near the boundary, we can study the scalar equations in the fixed geometric background by refraining from the backreactions \cite{Bianchi3}. The second is the arguments in \cite{deHaroSolodukhinSkenderis} that for the fields corresponding to the marginal operators, the leading backreaction is in $u^2$ order and so is negligible as $u\rightarrow 0$--Indeed by doing a Taylor's series expansion of the solution (\ref{eq12b}) around $u=0$, the first nontrivial-nonzero term is the second derivative one (proportional with $u^2$). Finally, If one looks at the associated charges and action corrections evaluated on the solutions (see, for instance (\ref{eq17}) and (\ref{eq22})), he/she clearly see that in the 11D supergravity legality limit, $N\gg k^5$, the object charges and their contributions to the actions are very small in the so-called probe approximations.}

\subsection{The Third Ansatz and Solution}
The third new ansatz we employ reads
\begin{equation}\label{eq19}
      G_7^{(4)}= idf_4 \wedge \mathcal{A}_3^{(0)} \wedge J \wedge e_7 + i f_4\ G_4^{(0)} J \wedge e_7 - i f_4\ \mathcal{A}_3^{(0)} \wedge J^2,
\end{equation}
which couples to an electric (anti)M5-brane wrapping some directions (according to the ansatz's structure) of the 11D space-- We discuss the reason why we call the associated objects M-branes or anti-M-branes at the first footnote in section \ref{sec.4}. From the Bianchi identity we obtain
\begin{equation}\label{eq20}
  \frac{d^2f_4(u)}{du^2} -\frac{2}{u}\frac{df_4(u)}{du}-\frac{1}{16u^2}f_4(u)=0,
\end{equation}
where, in obtaining the equation, we have used the conventions
\begin{equation}\label{eq20b}
  \ast_7 (J \wedge e_7)= 2 R\ J^2, \quad \ast_7 J^2=\frac{1}{2R} J \wedge e_7,
\end{equation}
with the solution
\begin{equation}\label{eq20c}
  f_4(u)=c_8\ u^{\frac{3}{2}-\frac{1}{4}\sqrt37} + c_9\ u^{\frac{3}{2}+\frac{1}{4}\sqrt37},
\end{equation}
as a pseudoscalar, in which the powers are actually $\Delta_- = 0 + \delta \Delta$, $\Delta_+ = 3 + \delta \Delta$ with $\delta \Delta=0.0273$, respectively. Although this solution is almost identical with the other ones here, an interesting point is that as approaching the boundary, $u \rightarrow 0$, the solution grows. This means that with $\Delta_+ \gtrsim 3$, one may propose a \emph{marginally irrelevant} deformation for the boundary field theory. Still, with $G_4 = G_4^{(0)}+ G_4^{(4)}$, the second term of the $\mathcal{A}_3$ field equation (\ref{eq04a}) is satisfied just with the condition
\begin{equation}\label{eq20d}
  \frac{df_4(u)}{du}-\frac{3}{u}
  f_4(u)=0 \Rightarrow f_4(u)=\tilde{c}_{9} u^3.
\end{equation}

The electric charge based on the last solution (\ref{eq20d}), in the unit internal 7D volume, is
\begin{equation}\label{eq21a}
 Q_e^4= - \frac{3 \tilde{c}_{9}}{64 \pi^6} \sqrt \frac{3 k^5}{2 R^{11}},
\end{equation}
with a 3-sphere of the radius $R/2$ to simulate three boundary coordinates, and $Vol(CP^3 \ltimes S^1/Z_k)=\frac{\pi^4 R^7}{3k}$; while the magnetic charge reads
\begin{equation}\label{eq21b}
 Q_m^4= - \frac{3 \tilde{c}_{9}}{16 \pi^7} \sqrt \frac{3 k^3}{2 R^{23}} \Lambda^3,
\end{equation}
with the upper limit $u_{\rightarrow\infty} = \Lambda$ ($=R/2$ may be taken) for the $u$ integration.

With the same conventions to evaluate the charges, the correction to the action from (\ref{eq03}), (\ref{eq19}) and (\ref{eq20d}), in the unit internal 7D volume, becomes
\begin{equation}\label{eq22}
  S_{M.4}^{E} = - \frac{\tilde{c}_9^2}{128 \pi^4} \sqrt \frac{3 k^3}{R^{11}} \Lambda^3.
\end{equation}
One may note to the reverse sign of the charges and action in the case, with respect to the original one. This means that we are allowed to interpret the case as anti-M5-branes wrapping some internal and external directions to produce the objects along the $u$ direction; indeed a string of the monopole-instantons are in some parallel planes orthogonal to the $u$ direction, say $u_{\rightarrow\infty} = \Lambda$.

For the backreaction issue, similar to the second solution, we note that
\begin{equation}\label{eq23}
       G_{MNPQ}^{(tot.4)}=G^{(0)}_{\acute{\mu}\acute{\nu}\acute{\rho}\acute{\sigma}} + G^{(4)}_{\acute{\mu} m n \varphi} = N \varepsilon_{\acute{\mu}\acute{\nu}\acute{\rho}\acute{\sigma}}- \frac{1}{12} \partial_{\acute{\mu}} f_4\ J_{m n},
\end{equation}
and so, one can check by (\ref{eq04b1}) that the corrections to the energy-momentum tensor, for both internal and external components, do not vanish and then, due to this backreaction, we cannot simply uplift the 4D solution to a complete 11D one.

\subsection{Symmetries and Duality}
We first note that from analyzing the solutions near boundary, as $f(u,\vec{u}) \approx \alpha(\vec{u})\ u^{\Delta_-} + \beta(\vec{u})\ u^{\Delta_+}$, see that we have indeed the massless (pseudo)scalars that couple to the conformal operators $\Delta_\mp=0,3$. The normalizable bulk mode is equivalent to the operator $\Delta_+=3$ and Dirichlet boundary condition $\delta\alpha=0$, where the source $\alpha$ couples to the boundary operator in this \emph{usual} CFT \cite{KlebanovWitten}. In the case, we have the following dictionary
\begin{equation}\label{eq24}
\begin{split}
& S_{on}[\alpha]=-W[\alpha], \quad \frac{1}{3} \langle \mathcal{O}_3(\vec{u})\rangle_\alpha =-\frac{\delta W[\alpha(\vec{u})]}{\delta\alpha(\vec{u})}=\beta(\vec{u}),\\
 & \ \ \ \ \ \ S \rightarrow S + W, \quad  W= -\frac{1}{3} \int d^3 \vec{u} \ \alpha(\vec{u})\ \mathcal{O}_3(\vec{u}),
\end{split}
\end{equation}
where $S_{on}$ and $W$ stand for the bulk on-shell action and boundary generating functional respectively, given that the source is a delta function $\delta^3(\vec{u}-\vec{u}_0)$ near the boundary.

The next point is about the conformal $SO(4,1)$ transformation $x_{\acute{\mu}} \rightarrow \frac{x_{\acute{\mu}}}{u^2+r^2}$ with $r=\sqrt{x_i x^i}$, which maps a point at infinity to origin. The compact space is now $S^3 \times S^7/Z_k$ and in general the associated brane-directions are reversed by the map. The background 4-form flux (\ref{eq02}), and the first ansatzs (\ref{eq05}, \ref{eq06}), just changes a sign under the transformation, which is indeed a skew-whiffing that changes M2-branes to anti-M2-branes. So, the solutions transform into one like (\ref{eq12b}). But, the second and third anstazs (\ref{eq11}, \ref{eq19}) are not invariant and do not change just a sign under the transformation. In other words, while the Laplacian (\ref{eq12}) is an invariant by the map, the equation (\ref{eq20}) is not and so the matching solution (\ref{eq20d}) does not give much information about the space structure. Meanwhile, the pseudoscalar solution from the third ansatz (\ref{eq19}), not corresponding to an \emph{exactly marginal} operator, may break the isometry of $EAdS_4$ slightly.

We now deal with isometry and supersymmetry issue of the solutions. The first ansatzs are indeed singlet under the original $SO(8)$ and $SU(4)$ isometry groups and are neutral for $U(1)$. The second and third ansatzs are also $SU(4)\times U(1)$ invariant in that both $J$ and $e_7$ are singlet under the latter group. Also, we note that after $Z_k$ orbifolding, by increasing $k$ and indeed with $k\geq 3$, the isometry group becomes that of $CP^3$ and the supersymmetry becomes $\mathcal{N}=6$. Then, for $k=1,2$ the internal space isometry is increased to $SO(8)$ and supersymmetry to $\mathcal{N}=8$ due to monopole operators \cite{GustavssonRey} and so, our dual solutions should also respect them.

The main supersymmetry is preserved when we add M2-branes in the same directions with the original ones. But by the skew-whiffing or parity transforming on the original solution, or equivalently introducing the anti-M2-branes associated with the conformally transforming the ansatzs (\ref{eq05}, \ref{eq06}), the supersymmetry is completely broken by the scalars. The second and third solutions also break all supersymmetries. Indeed, for the second ansatz (\ref{eq11}), the number of \emph{relative transverse directions} between the associated (anti)M2(M5)-brane and the original M2-branes is 2(6)-- known as non-threshold BPS bound-states in \cite{9612095}-- and so, the pseudoscalar there breaks all supersymmetries. The third ansatz (\ref{eq19}), the 7-form field, contains three terms and the associated (anti)M5-brane can indeed wrap around various directions depended on the terms selected in the $\omega$ and $J$ and so, its pseudoscalar breaks all supersymmetries as well, and the same for its electric dual (anti)M2-brane.

Therefore, based on the symmetries and the objects we found in the bulk, we should adjust or deform the dual filed theory to find the best matching operators and solutions, as we do in the next section with three boundary duals.

\section{On Conformal-Field Theory Side}
\subsection{Dual Scalars}\label{sub3.1}
From the first ansatzs, we have real scalars in the bulk of $EAdS_4$ because of the form field $\mathcal{A}_{xyz}$ and so, on the boundary theory we expect to have the same objects. On the other hand, we remember that the massless scalars are in the representation $\textbf{35}_{v} = \textbf{15}_{0} \oplus \textbf{10}_{2} \oplus \bar{\textbf{10}}_{-2}$ of $SO(8) \rightarrow SU(4) \times U(1)$ from the original theory \cite{ABJM}, and the pseudoscalars are in $\textbf{35}_c = \textbf{15}_{0} \oplus \textbf{10}_{-2} \oplus \bar{\textbf{10}}_{2}$ representation \cite{Halyo}. That is when the gravitons, fermions and scalars are in the representations $\textbf{8}_s = \textbf{1}_{2} \oplus \textbf{1}_{-2} \oplus \textbf{6}_{0}$, $\textbf{8}_c=\textbf{4}_{-1} \oplus \textbf{4}_{1}$ and $\textbf{8}_v= \textbf{4}_{1} \oplus \bar{\textbf{4}}_{-1}$, respectively. In the case we have $\mathcal{N}=8$ supersymmetry which in turn breaks into $\mathcal{N}=6$ when considering $S^7/Z_k$ as a $S^1/Z_k$ fiberation on $CP^3$ for a generic $k\neq 1,2$.

Now, for the scalars from the M2-branes in the same directions with the original ones and the solution like (\ref{eq06b}), the best $\Delta_+=3$ $SU(4)\times U(1)$-singlet (and also $SO(8)$ invariant \cite{GustavssonRey}) agreeing operator has the same structure with the scalar potential of the model, which could simply be $\hat{\mathcal{O}}_3= tr\big(Y^A Y_A^\dagger Y^B Y_B^\dagger Y^C Y_C^\dagger\big)$, where $Y^A$ ($A=1,2,3,4$) are the four complex scalars of the ABJM model that transform in the bifundamental representation $\textbf{4}_1$ of $SU(4)_R \times U(1)_b$, with the $b$ index for the baryonic symmetry. Now with $Y^A=Y_A^\dagger= \frac{b_1}{N} \emph{\textbf{I}}_{N \times N}$, where $\emph{\textbf{I}}_{N \times N}$ is the unitary matrix and $b_1, b_2,...$ are some boundary constants, for the real boundary scalars we have the trivial result
\begin{equation}\label{eq28}
     \langle \mathcal{\hat{O}}_3 \rangle_{c_3}=64 b_1=3 c_4.
\end{equation}

On the other hand, for the conformally transformed (skew-whiffed) case, where the solution is like (\ref{eq12b}), the supersymmetry is broken completely. This situation is handled by an exchange of the representations as $\textbf{8}_s \rightarrow \textbf{8}_v$ and so the scalars now set in $\textbf{35}_{v} \rightarrow \textbf{35}_{s}=\textbf{1}_{0} \oplus \bar{\textbf{1}}_{4} \oplus \textbf{1}_{-4} \oplus \bar{\textbf{6}}_{2} \oplus \textbf{6}_{-2}\oplus \bar{\textbf{20}}_{0}$, while the pseudoscalars remain in $\textbf{35}_{c}$. In the case we have the wished singlet $\textbf{1}_{0}$ scalar in the spectrum. In other words, we can break up the eight scalars as $X^I \rightarrow (y^n, y, \bar{y})$ with  $I,J...=(1,...6,7,8)=(n,7,8)$, $y=y^7+iy^8$ and $y^\dagger=\bar{y}$. Then one can simply construct the desired operator. Starting from a rank-2 symmetric traceless BPS operator composed of the scalars $X^I$, we arrive at the operator
\begin{equation}\label{eq29}
    \mathcal{O}_1= \frac{3}{4} tr(y \bar{y}),
\end{equation}
which is singlet under $SO(8)$ and $SU(4)\times U(1)$--We also note that one could consider the singlet dimension -1 non-BPS Konishi operator $\hat{\mathcal{O}}_1=tr(Y^A Y_A^\dagger)$ to do the same job.

Next, if we make the antisymmetrized expansion of the bosonic potential, which has the operator structure
\begin{equation}\label{eq29}
    \mathcal{O}_{3a} = tr\big(X^{[I} X_J^\dagger X^{K]} X_{[K}^\dagger X^J X_{I]}^\dagger \big),
\end{equation}
and set $y^n=0$, see that it vanishes. Further, by setting the gauge fields $A_i$ and $\hat{A}_i$ and the fermion fields $\psi_A$ to zero, and deforming the remaining part of the original Lagrangian, according to (\ref{eq24}) with $\mathcal{O}_{3a}=\mathcal{O}_1^3$, we have
\begin{equation}\label{eq30}
     \mathcal{L}_{def.}=-tr(\partial_i y \partial^i \bar{y}) - \frac{9 c_3}{64} tr (y\bar{y})^3,
\end{equation}
for the skew-whiffed solution of (\ref{eq06b}). Then, from the EOM of the Lagrangian and setting
\begin{equation}\label{eq30b}
     y= i \frac{h(r)}{N} \emph{\textbf{I}}_{N \times N},
\end{equation}
where the convention for the matrix-valued fields in the large $N$ limit are form \cite{Witten2}, we get
\begin{equation}\label{eq30c}
    h(r)= \sqrt[1/2]{\frac{64 N^2}{27 c_3}} \bigg(\frac{b_2}{b_2^2+r^2}\bigg)^{1/2},
\end{equation}
and the value of the action based on the solution reads
\begin{equation}\label{eq30d}
   S_{def.1}= -\frac{8}{9} \sqrt{\frac{4}{3 c_{3}}}\ \pi^2,
\end{equation}
where we have used the integrations
\begin{equation}\label{eq31}
   \int_0^\infty \frac{b_2\ r^4}{(b_2^2 + r^2)^3} dr=\frac{3 \pi}{16}, \qquad  \int_0^\infty  \frac{b_2^3\ r^2}{(b_2^2 + r^2)^3} dr=\frac{\pi}{16}.
\end{equation}
Now, as a preliminary benchmark, one can simply see that the one-point function of the operator $\mathcal{O}_{3a}$ for the solution (\ref{eq30c}) fits with $\beta \sim \frac{c_4}{r^6}$ from the skew-whiffing of (\ref{eq06b}), according to (\ref{eq24}); so the dual constants adjust as well.

\subsection{Agreeing Gauge fields}
The gauge fields in all gravitino representations transform as $\textbf{28} \rightarrow \textbf{1}_{0} \oplus \bar{\textbf{6}}_{2} \oplus \textbf{6}_{-2} \oplus \textbf{15}_{0}$ when $SO(8)\rightarrow SU(4) \times U(1)$. On the other hand, we note that for $k=1, 2$ the latter symmetry with $\mathcal{N}=6$ is enhanced to the former with $\mathcal{N}=8$, which is the case valid with two M2-branes and the gauge group $SU(2) \times SU(2)$ of BLG as well \cite{1203.3546}. In addition, one should also note that although the gauge fields and the scalar potentials are universal in both ABJM and BLG models \cite{Raamsdonk}, \cite{GustavssonRey}, to fit the bulk pseudoscalars on the boundary, the best way is to use the boundary gauge fields or deforming the original theory by fermions that we made a sample in \cite{Me3} and make another in the next subsection \ref{sub3.3}.

Meanwhile, we remind that the second and third ansatzs (\ref{eq11}, \ref{eq19}) do not preserve conformal symmetry (violate parity invariance), or according to the novel Higgs mechanism \cite{ChuNastaseNilssonPapageorgakis}, related to the discussion in the previous subsection (\ref{sub3.1}) too, the original quiver gauge group breaks down into one $SU(N)$ \cite{ABJM}. Another interesting reason here is that at least for the third ansatz, it seems that we may consider the associated (anti)M5-brane, whose world-volume wraps around three internal directions (like $S^3/Z_k$ as in ABJ) and the other three are parallel to those of the original M2-branes, as \emph{fractional} (anti)M2-branes \cite{ABJ}. In the latter case with two fractional (anti)M2-branes, the gauge group $SU(N+2)_k \times SU(N)_{-k}$ could be broken into $SU(N)_k \times SU(N)_{-k} \times SU(2)$ and then, one could keep the main quiver group as spectator and consider the $SU(2)$ part in action.

Anyway, we set all scalars, fermions and the gauge field $\hat{A}_i$ to zero, in the main M2-brane action--see for instance \cite{Terashima} and \cite{Me3, Me4} for the notation--and so
\begin{equation}\label{eq32}
   \mathcal{L}_{CS} = \frac{ik}{4\pi}\, \varepsilon^{ijk} \, tr \bigg(A_i \partial_j A_k+\frac{2i}{3} A_i A_j A_k \bigg),
\end{equation}
with $i$ factor because of being in the Euclidean space, and that $F_{ij}=0$ here so that $A_i$ behaves like a pure gauge at infinity. To this end, we use the singular gauge
\begin{equation}\label{eq33}
   A_i=\frac{u^2}{u^2+r^2} g^{-1} \partial_i g, \quad g = \frac{(u {1}_2 - i x_i \sigma^i)}{\sqrt{u^2 + r^2}},
\end{equation}
with $g$ as an element of $SU(2)$ for our purpose. Therefore, the value of the action reads $S_{def.2}=2 k  \pi$, where we have used
\begin{equation}\label{eq34}
   \int_{S_\infty^3}\ tr\big(A^3\big)=24 \pi^2, \quad F=dA + i A \wedge A,
\end{equation}
with a note that the pseudoscalar solutions of the Yang-Mills equations were first studied in \cite{Belavin}. Now, the interesting point is that the singlet dimension-3 operator has a structure like
\begin{equation}\label{eq35}
  \mathcal{O}_{3b} \sim tr (A \wedge F) \sim \frac{u^3}{(u^2+r^2)^3},
\end{equation}
whose expectation value next to the boundary is $\langle \mathcal{O}_{3b} \rangle_{c_5} \sim \frac{c_6}{r^6}$ with respect to the bulk solution (\ref{eq12b}) and the standard dictionary (\ref{eq24})--For more details look at \cite{Me4}, and also \cite{Kogan} for a similar study on D-instantons in type IIB supergravity over $AdS_5 \times S^5$.

\subsection{Dual Pseudoscalars} \label{sub3.3}
Beside the gauge fields to comply with the bulk pseudoscalars, one may search for boundary pseudoscalars to form the dimension-3 $SU(4) \times U(1)$-singlet operator. To handle this issue, we note that the pseudoscalars are in the $\textbf{35}_c = \textbf{15}_{0} \oplus \textbf{10}_{-2} \oplus \bar{\textbf{10}}_{2}$ representation of the original theory. But, by skew-whiffing $\textbf{8}_s \rightarrow \textbf{8}_c$, while the scalars remain in $\textbf{8}_v$ unchanged, the pseudoscalars sit in $\textbf{35}_{c} \rightarrow \textbf{35}_{s}=\textbf{1}_{0} \oplus \bar{\textbf{1}}_{4} \oplus \textbf{1}_{-4} \oplus \bar{\textbf{6}}_{2} \oplus \textbf{6}_{-2}\oplus \bar{\textbf{20}}_{0}$ including the wanted singlet bulk pseudoscalars, $f_3$ and $f_4$ from the second and third ansatzs, and attention that the supersymmetry is broken completely.

Then, for the suiting boundary operator, we first used $\tilde{\mathcal{O}}_3= tr \big(Y_A^\dagger Y^A \psi^{B\dagger} \psi_B \big)$ with $\psi_A$ for the fermions transforming as $\bar{\textbf{4}}_{-1}$ of $SU(4)_R \times U(1)_b$ \cite{Me3}. \footnote{Why we are considering these special boundary operators to adjust the bulk states with, comes mainly from the precession spectroscopy of the involved theories; look for instance at \cite{Bianchi2} and \cite{Forcella.Zaffaroni}. In addition, scalars in the bulk should usually be adjusted with the boundary operators formed from the boundary scalars. The bulk pseudoscalars could be adjusted with the gauge fields (pseudo-particles as stated in \cite{Belavin}) and, according to resultant boundary theory and its broken symmetries, fermions can also come into the play.} Now, we take $\psi$ as the singlet $\textbf{1}_{2}$ spinor field in the new $\textbf{8}_c$ and
\begin{equation}\label{eq36}
  \mathcal{O}_{3c}=\big(tr(\psi \bar{\psi})\big)^{3/2},
\end{equation}
as the suitable singlet $\Delta_+=3$ operator. Next, one can see that to get a matching solution from the field equations, we need to also keep the gauge fields while set the scalars to zero. So, the deformed action, according to (\ref{eq24}), reads
\begin{equation}\label{eq37}
   S_{def.3} = \int d^3\vec{u}\  tr \bigg\{\mathcal{L}_{CS} + \hat{\mathcal{L}}_{CS}- tr \big(\psi^{\dagger} i \gamma^i D_i \psi \big)- \frac{c_5}{3} \big(tr(\psi \bar{\psi})\big)^{3/2} \bigg\},
\end{equation}
where $\hat{\mathcal{L}}_{CS}$ is that in (\ref{eq32}) with exchanging $A$ with $\hat{A}$, and note that $D_i \psi =\partial_i \psi + i A_i \psi - i \psi \hat{A}_i$. The field equations so become
\begin{equation}\label{eq38a}
     i \gamma^i D_i \psi -\frac{c_5}{2} \big(tr(\psi \bar{\psi})\big)^{1/2} \psi=0,
\end{equation}
\begin{equation}\label{eq38b}
     \frac{ik}{4\pi} \varepsilon^{ijk} F_{jk}+ \bar{\psi} \gamma^i \psi=0, \quad \frac{ik}{4\pi} \varepsilon^{ijk} \hat{F}_{jk}+ \bar{\psi} \gamma^i \psi=0.
\end{equation}
Then, to find a clear solution, we set $\psi_{\hat{a}}^a= \frac{\delta_{\hat{a}}^a}{N} \psi$ that is equivalent to concentrating on the $U(1) \times U(1)$ part of the full gauge group. On the other hand, by setting $A_i^\pm \equiv (A_i \pm \hat{A}_i)$, one can write
\begin{equation}\label{eq38c}
      \frac{i k}{4 \pi} \varepsilon^{ijk} F_{jk}^+ = -2 \bar{\psi} \gamma^i \psi, \quad F_{jk}^- = 0,
\end{equation}
with a note that the ABJM matter fields are neutral to $A_i^+$ while $A_i^-$ works as the baryonic symmetry and so, setting $A^-=0$ means searching for a self-interacting spinor field.

Now, by taking a similar ansatz as that in \cite{I.N} and \cite{Me3} for solving the $\psi$ equation, we arrive at
\begin{equation}\label{eq39}
      \psi= \frac{6 b_3 \sqrt{N}}{c_5} \frac{\big(b_3 + i (x-x_0)^i \gamma_i \big)}{\big(b_3^2 + (x-x_0)_i (x-x_0)^i \big)^{3/2}} \bigg(\begin{array}{c}   1  \\     0   \end{array}\bigg),
\end{equation}
where the use is made of $\gamma_i=(\sigma_2, \sigma_1, \sigma_3)$ as the Euclidean gamma matrices. The finite value of the action (\ref{eq37}) on the solution (\ref{eq39}), and with help from (\ref{eq38a}), now reads
\begin{equation}\label{eq41}
   S_{def.3} = \frac{c_5}{6} \int d^3\vec{u}\ \big(tr(\psi \bar{\psi})\big)^{3/2} =\frac{36\ b_3^2}{c_5^2} \int_0^\infty \frac{d^3\vec{u}}{\big(b_3^2+(\vec{u}-\vec{u}_0)^2\big)^3}= \frac{9 \pi^2}{c_5^2\ b_3},
\end{equation}
that signals the instanton nature of the solution.

Meanwhile, to evaluate the net magnetic charge for the $A^+$ field, we write
\begin{equation}\label{eq40}
     B^i=\frac{1}{2} \varepsilon^{ijk} F^+_{jk}=\frac{48 \pi i b_3^2 N}{k c_5^2 (b_3^2 + r^2)^3} (2 x b_3, -2y b_3, -(b_3^2 + r^2)) \Rightarrow \Phi=\oint_{\hat{s}} \vec{B}. d\vec{s}=0,
\end{equation}
with $\hat{s}$ for $S^3_\infty$, pointing out that the solution (\ref{eq39}) in indeed $SU(4)_R \times U(1)_b$-singlet.

The primary test for the correctness of the dual solution comes from
\begin{equation}\label{eq41}
     \langle \mathcal{O}_{3c} \rangle_{c_5} \sim {\big(b_3^2+(\vec{u}-\vec{u}_0)^2\big)^{-3}},
\end{equation}
which is proportional with $\beta({\vec{u}})$ according to (\ref{eq12b}) and (\ref{eq24}) and so, one can simply connect the constants $c_5, b_3$, and $c_6$ together.

An interesting point is that the single-trace deformation here is equivalent to the multi-trace deformation in \cite{I.N} with similar solutions. In fact, there we found a conformally coupled $m^2=-2$ pseudoscalar in the bulk of Euclidean $AdS_4$, where the resultant equation was cubic in the field and conformally equivalent to a massless pseudoscalar in the flat space. A sign of agreement comes from the used ansatzs, where the third ansatz here, associated with a new included (anti)M2-brane, has almost the same structure $\mathcal{A}_3 \sim (f J \wedge e_7)$ as the predecessors. As a result, one may connect the parameters of both sides and do parallel discussions on some aspects such as supersymmetry breaking by the special multi-trace deformations and instability because of the solution, as discussed in \cite{deHaro2} as well.

\section{Discussions and Comments} \label{sec.4}
In this paper, we have introduced some form fields coupled to (anti)M2- and (anti)M5-branes of 11-dimensional supergravity over $AdS_4 \times S^7/Z_k$. From them we earned some massless scalars and pseudoscalars in the bulk and tried to find the counterpart dimension-3 operators on the boundary from the scalars, fermions and gauge fields of the standard M2-branes Lagrangian. For the bulk scalars, to avoid the backreactions, we made a special parity transformation equivalent to adding some anti-M2-branes to the directions-reversed (skew-whiffed) background M2-branes or vice versa, and then the dual $SO(8)$ and $SU(4) \times U(1)$ singlet scalars were found from the sextic scalar potential of the Lagrangian. The bulk pseudoscalars, associated with some special (anti)M2- and (anti)M5-branes, \footnote{To clarify more and hint the point why we mention both branes and anti-branes, we note that from the second and third ansatzs we have indeed \emph{singlet} (under the internal isometry group) \emph{pseudoscalars} in the bulk. Therefore, we may think of the ansatzs as associated with some probe anti-M-branes added on the original (M2-branes) background or as some probe M-branes added to the skew-whiffed (anti-M2-branes) background and then, the resultant theory is for anti-M2-branes. As a result, and to explain the bulk states, we are forced (indeed the best way is) to do skew-whiffing (fulfilled by some swaping of the $SO(8) \rightarrow SU(4) \times U(1)$ group representations) on the boundary with breaking all supersymmetries.}\label{fotnot6} backreact on the geometry and their dual boundary solutions were made from the agreeing $SU(2)$ gauge fields and a special deformation by a singlet fermion field of anti-M2-branes (skew-whiffed) theory. In the following lines we comment on some other related issues.

The second and third ansatzs have a few interesting features. From the second ansatz (\ref{eq11}), we have an (anti)M5-brane wrapping all internal $CP^3$ space resulting in a fully localized object in the bulk of $EAdS_4$ that we call it \emph{instanton}; while its dual electric (anti)M2-brane has two tangent directions on the external space and may be associated with a string. For the (anti)M5-brane from the third ansatz (\ref{eq19}), one can immediately see that it can have three tangent directions on the external space and so, the solutions may be considered as \emph{domain walls}; while its dual electric (anti)M2-brane could wrap around $CP^1 \times S^1/Z_k$ resulting in a string of \emph{monopole-instantons} along the $u$ direction.

Another point is on supersymmetry breaking by the solutions. Although for the first ansatz the skew-whiffing breaks all supersymmetries expect for $S^7$, the second and third solutions break all supersymmetries originally in that the associated M-branes wrap around some mixed internal directions. But, one can still consider some special internal directions for the M-branes to be wrapped on with preserving some supersymmetries. A well-known example with Minkowskian signature, related to the third ansatz, is the so-called \emph{fuzzy 3-sphere funnel type solutions}. Indeed, in \cite{Bena} the vacua of a probe M5-brane with the topology $R^{1,2} \times S^3$, which has three directions in common with the background M2-branes, are investigated. There are domain walls interpolating among the vacua in the model. Next, it is shown in \cite{Terashima} that, from BPS and non-BPS equations of the ABJM action, one can have equations equivalent to \emph{Basu-Harvey equations} describing bound-states of M2-M5 brane systems with preserving some or breaking all supersymmetries \footnote{An earlier and similar study with BLG model is done in \cite{Krishnan1}.}. These points are further investigated in \cite{HanakiLin}, where the Basu-Harvey type equations describe $N$ M2-branes ending on a M5-brane wrapping a fuzzy $S^3/Z_k$. It is also shown that the \emph{Nahm equation}, describing $N$ D2-branes ending on a D4-brane wrapping a $S^2 \sim CP^1$, comes from the large $k$ limit of the former equations. This fact is also related to the idea of the fractional M2/D2-branes presented in the ABJ study \cite{ABJ}.

The instability of the solutions is verified by a few reasons. First, one should note that the marginal operators corresponding to the massless bulk fields and nonsupersymmetric solutions destabilize the vacua although in general quantum corrections should be taken into account. Forming brane-antibrane pairs, and multi-trace deformations destabilize the vacua \cite{Witten2}, \cite{deHaro2} as well. In addition, and more relevant to the case, it is argued in \cite{Narayan} that nonsupersymmetric Chern-Simons theories with $C^4/Z_k$ orbifold singularities are unstable, especially when one considers the quotient space as a $S^1$ Kaluza-Klein compactification on the weighted $CP^3$. In other words, according to \cite{Murugan}, the \emph{global singlet marginal operators} can disturb the conformal fixed points in nonsupersymmetric theories when there are  of course orbifold spaces or some skew-whiffing, which is valid here too. Then, the tunneling decay of these spaces into \emph{a bubble of nothing} \cite{Witten3} is because of a shrinking circle. This is mainly a case with our second anstaz, $G_4^{(3)}$  of (\ref{eq11}), where the fiber coordinate is indeed for the KK $S^1/Z_k$ that wrapping around it causes nonsupersymmetric instability. It is also mentionable that from (\ref{eq15}), one clearly see that for $k=4$ the associated electric charge vanishes and for $k>4$ its sign changes, signaling instability.

Another remarkable point is that as we know that 10D type IIA supergravity is always considered as a dimensional reduction of the 11D supergravity and so, one expects that all solutions in the latter case have counterparts for the former. Indeed, it is known that D2- and NS5-branes on one side and F1-string and D4-branes on the other side come from direct- and double-dimensional reductions of M2- and M5-branes, respectively \cite{PKTownsend01}, \cite{Bergshoeff02} (next to the novel Higgs mechanism surveyed in \cite{ChuNastaseNilssonPapageorgakis} to relate M2- and D2-branes). Therefore, checking the issue with the ansatzs here, and whether one may learn something about the 11D supermembrane theory from them, will be interesting. On the other hand, it will be fascinating to investigate instability of M-branes (having continuous spectrums) starting from the original work in \cite{DEWIT01} using their dual formulation in terms of D-branes-- An interesting work in this phase for the ABJM model is done in \cite{1102.3277}.

Finally, in addition to the references quoted sofar on (mainly nonperturbative) studies in framework of AdS$_4$/CFT$_3$ correspondence, here is a good place to mention some other related and recent works. For a study on various partially localized objects (monopoles, vortices, Q-balls and so on) in the ABJM model, look at \cite{1211.5886} and references therein, and see \cite{1408.0912} for studying the type IIB T-dual of the extended objects in the correspondence.
To study localization techniques in supergravity in the current duality and context look at \cite{1406.0505}, and see \cite{1406.1892} for a study of adding M5-branes in $AdS_4 \times Q^{1,1,1}$ space-times and also look at \cite{1411.6170} for an overview of exact (intersecting) solutions in M-theory. Some nonperturbative effects in M-theory (including Fermi gas approach and instanton corrections) are discussed in \cite{1407.3786} and references therein and similar exact results in $\mathcal{N}=8$ Chern-Simon-matter theories are found in \cite{1409.1799} based on localization techniques. The investigating of Wilson loops in ABJM model are done in \cite{1507.00442} and references therein, and a recent study on flux, conductivity and (fractional) quantum Hall states in the model is done in \cite{1411.3335}.

\section{Acknowledgement}
This work is supported by a Grant from the University of Ilam.


\begin{thebibliography}{99}
\bibitem{Duff84NPW} M. J. Duff, B. E.W. Nilsson, C. N. Pope and N. P. Warner, \textit{"On the consistency of the Kaluza-Klein
ansatz"}, \href{http://www.sciencedirect.com/science/article/pii/0370269384915582}{Phys. Lett. B 149, 90 (1984)}.
\bibitem{Gauntlett} J. P. Gauntlett and O. Varela, \textit{"Consistent Kaluza-Klein reductions for general supersymmetric AdS solutions"}, Phys. Rev. D 76, 126007 (2007), \href{http://arxiv.org/abs/0707.2315}{[arXiv:0707.2315 [hep-th]]}.
\bibitem{ABJM} O. Aharony, O. Bergman, D. L. Jafferis and J. Maldacena, \textit{"$\mathcal{N}$=6 superconformal Chern-Simons matter theories, M2-branes and their gravity duals"}, JHEP 0810, 091 (2008), \href{http://arxiv.org/abs/0806.1218}{[arXiv:0806.1218 [hep-th]]}.
 \bibitem{1203.3546} J. Bagger, N. Lambert, S. Mukhi and C. Papageorgakis, \textit{"Multiple membranes in M-theory"}, Phys. Rept. 5271,  (2013), \href{http://arxiv.org/abs/1203.3546}{[arXiv:1203.3546 [hep-th]]}.
\bibitem{I.N} A. Imaanpur and M. Naghdi, \textit{"Dual Instantons in Anti-membranes theory"}, Phys. Rev. D 83, 085025 (2011), \href{http://arxiv.org/abs/1012.2554}{[arXiv:1012.2554 [hep-th]]}.
\bibitem{N} M. Naghdi, \textit{"A monopole Instanton-like effect in the ABJM model"}, Int. J. Mod. Phys. A 26, 3259 (2011), \href{http://arxiv.org/abs/1106.0907}{[arXiv:1106.0907 [hep-th]]}.
\bibitem{I} A. Imaanpur, \textit{"U(1) Instantons on $AdS_4$ and the uplift to exact supergravity solutions"}, JHEP 1111, 041 (2011), \href{http://arxiv.org/abs/1108.2786}{[arXiv:1108.2786 [hep-th]]}.
\bibitem{Me3} M. Naghdi, \textit{"New Instantons in AdS$_4$/CFT$_3$ from D4-Branes Wrapping Some of CP$^3$"}, Phys. Rev. D 88, 026013 (2013),
    \href{http://arxiv.org/abs/1302.5294}{[arXiv:1302.5294 [hep-th]]}.
\bibitem{Me4} M. Naghdi, \textit{"Marginal Fluctuations as Instantons on M2/D2-Branes"}, Eur. Phys. J. C 74, 2826 (2014), \href{http://arxiv.org/abs/1302.5534}{[arXiv:1302.5534 [hep-th]]}.
\bibitem{KlebanovWitten} I. R. Klebanov and E. Witten, \textit{"AdS/CFT correspondence and symmetry breaking"}, Nucl. Phys. B 556, 89 (1999), \href{http://arxiv.org/abs/hep-th/9905104}{[arXiv:hep-th/9905104]}.
\bibitem{Park} C. Park and Sand-J Sin, \textit{"Notes on D-instantons cortrections to $AdS_5 \times S^5$ geometry"}, Phys. Lett. B 444, 156 (1998), \href{http://arxiv.org/abs/hep-th/9807156}{[arXiv:hep-th/9807156]}.
\bibitem{Bianchi3} M. Bianchi, D. Z. Freedman and K. Skenderis,\textit{"Holographic renormalization"}, Nucl. Phys. B 631, 159 (2002), \href{http://arxiv.org/abs/hep-th/0112119}{[arXiv:hep-th/0112119]}.
\bibitem{deHaroSolodukhinSkenderis} S. de Haro, S. N. Solodukhin and K. Skenderis, \textit{"Holographic reconstruction of spacetime and renormalization in the AdS/CFT correspondence"}, Commun. Math. Phys. 217, 595 (2001), \href{http://arxiv.org/abs/hep-th/0002230}{[arXiv:hep-th/0002230]}.
\bibitem{GustavssonRey} A. Gustavsson and S. J. Rey, \textit{"Enhanced $\mathcal{N}$=8 Supersymmetry of ABJM Theory on $R^8$ and $R^8/Z_2$"}, \href{http://arxiv.org/abs/0906.3568}{[arXiv:0906.3568 [hep-th]]}.
\bibitem{9612095} E. Bergshoeff, M. de Roo, E. Eyras, B. Janssen and J. P. van der Schaar, \textit{"Multiple intersections
of D-branes and M-branes,"}, Nucl. Phys. B 494, 119 (1997), \href{http://arxiv.org/abs/hep-th/9612095}{[arXiv:hep-th/9612095]}.
\bibitem{Halyo} E. Halyo, \textit{"Supergravity on $AdS_{5/4} \times$ Hopf fibrations and conformal field theories"}, Mod. Phys. Lett. A 15, 397 (2000), \href{http://arxiv.org/abs/hep-th/9803193}{[arXiv:hep-th/9803193]}.
\bibitem{Witten2} E. Witten, \textit{"Multi-trace operators, boundary conditions, and AdS/CFT correspondence"}, \href{http://arxiv.org/abs/hep-th/0112258}{[arXiv:hep-th/0112258]}.
\bibitem{Raamsdonk} M. Van Raamsdonk, \textit{"Comments on the Bagger-Lambert theory and multiple M2-brane"}, JHEP 0805, 105 (2008), \href{http://arxiv.org/abs/0803.3803}{[arXiv:0803.3803 [hep-th]]}.
\bibitem{ChuNastaseNilssonPapageorgakis} X. Chu, H. Nastase, B. Nilsson and C. Papageorgakis, \textit{"Higgsing M2 to D2 with gravity: $\mathcal{N}$=6 chiral supergravity from topologically gauged ABJM theory,"}, JHEP 1104, 040 (2011), \href{http://arxiv.org/abs/1012.5969}{[arXiv:1012.5969 [hep-th]]}.
\bibitem{ABJ} O. Aharony, O. Bergman, D. L. Jafferis, \textit{"Fractional M2-branes"}, JHEP 0811, 043 (2008), \href{http://arxiv.org/abs/0807.4924}{[arXiv:0807.4924 [hep-th]]}.
\bibitem{Terashima} S. Terashima, \textit{"On M5-branes in $\mathcal{N}=6$ membrane action"}, JHEP 0808, 080 (2008), \href{http://arxiv.org/abs/0807.0197}{[arXiv:0807.0197 [hep-th]]}.
\bibitem{Belavin} A. A. Belavin, A. M. Polyakov, A. S. Shvarts and Yu. S. Tyupkin, \textit{"Pseudoparticle solutions of the Yang-Mills equations"}, \href{http://www.sciencedirect.com/science/article/pii/037026937590163X}{Phys. Lett. B 59, 85 (1975)}.
\bibitem{Kogan} I. I. Kogan and G. Luz\'{o}n, \textit{"D-instantons on the boundary"}, Nucl. Phys. B 539, 121 (1999), \href{http://arxiv.org/abs/hep-th/9806197}{[arXiv:hep-th/9806197]}.
\bibitem{Bianchi2} M. Bianchi, R. Poghossian and M. Samsonyan, \textit{"Precision spectroscopy and higher spin symmetry in the ABJM model"}, JHEP 1010, 021 (2010), \href{http://arxiv.org/abs/1005.5307}{[arXiv:1005.5307 [hep-th]]}.
\bibitem{Forcella.Zaffaroni} D. Forcella and A. Zaffaroni, \textit{"Non-supersymmetric CS-matter theories with known AdS duals"}, Adv. High Energy Phys. 393645 (2011), \href{http://arxiv.org/abs/1103.0648}{[arXiv:1103.0648 [hep-th]]}.
\bibitem{deHaro2} S. de Haro, I. Papadimitriou and A. C. Petkou, \textit{"Conformally coupled scalars, instantons and vacuum instability in $AdS_4$"}, Phys. Rev. Lett. 98, 231601 (2007), \href{http://arxiv.org/abs/hep-th/0611315}{[arXiv:hep-th/0611315]}.
\bibitem{Bena} I. Bena, \textit{"The M theory dual of a three-dimensional theory with reduced supersymmetry"}, Phys. Rev. D 62, 126006 (2000), \href{http://arxiv.org/abs/hep-th/000414}{[arXiv:hep-th/000414]}.
\bibitem{Krishnan1} C. Krishnan and C. Maccaferri, \textit{"Membranes on calibrations"}, JHEP 0807, 005 (2008), \href{http://arxiv.org/abs/0805.3125}{[arXiv:0805.3125 [hep-th]]}.
\bibitem{HanakiLin} K. Hanaki and H. Lin, \textit{"M2-M5 systems in $\mathcal{N}$=6 Chern-Simons theory"}, JHEP 0809, 067 (2008), \href{http://arxiv.org/abs/0807.2074}{[arXiv:0807.2074 [hep-th]]}.
\bibitem{Narayan} K. Narayan, \textit{"On nonsupersymmetric $C^4/Z_N$, tachyons, terminal singularities and flips"}, JHEP 1003, 019 (2010), \href{http://arxiv.org/abs/0912.3374}{[arXiv:0912.3374 [hep-th]]}.
\bibitem{Murugan} A. Murugan \textit{"Renormalisation group flows in gauge-gravity duality"}, \href{https://www.princeton.edu/physics/graduate-program/theses/theses-from-2009/}{PhD Thesis, Princeton University (2009)}.
\bibitem{Witten3} E. Witten, \textit{"Instability of the Kaluza-Klein vacuum"}, \href{http://www.sciencedirect.com/science/article/pii/0550321382900074}{Nucl. Phys. B 195, 481 (1982)}.
\bibitem{PKTownsend01} P. K. Townsend, \textit{"D-branes from M-branes"}, Phys. Lett. B 373, 68 (1996), \href{http://arxiv.org/abs/hep-th/9512062}{[arXiv:hep-th/9512062]}.
\bibitem{Bergshoeff02} E. Bergshoeff, M. de Roo, E. Eyras, B. Janssen and J. P. van der Schaar, \textit{"Intersections involving waves and monopoles in eleven dimensions"}, Class. Quant. Grav. 14, 2757 (1997), \href{http://arxiv.org/abs/hep-th/9704120}{[arXiv:hep-th/9704120]}.
\bibitem{DEWIT01} B. de Wit, M. Luscher and H. Nicolai, \textit{"The supermembrane is unstable"}, \href{http://www.sciencedirect.com/science/article/pii/0550321389902149}{Nucl. Phys. B 320, 135 (1989)}.
\bibitem{1102.3277} T. Kuroki, A. Miwa and S. Okuda, \textit{"Deformation of half-BPS solution in ABJM model and instability of supermembrane"}, JHEP 1105, 011 (2011), \href{http://arxiv.org/abs/1102.3277}{[arXiv:1102.3277 [hep-th]]}.
\bibitem{1211.5886} G. Go, Ch. Kim, Y. Kim, O. K. Kwon and H. Nakajima, \textit{"BPS vortices, Q-balls, and Q-vortices in $\mathcal{N}$=6 Chern-Simons matter theory"}, JHEP 1302, 058 (2013), \href{http://arxiv.org/abs/1211.5886}{[arXiv:1211.5886 [hep-th]]}.
\bibitem{1408.0912} Y. Lozano and N. T. Macpherson, \textit{"A new $AdS_4/CFT_3$ dual with extended SUSY and a spectral flow"}, JHEP 1411, 115 (2014), \href{http://arxiv.org/abs/1408.0912}{[arXiv:1408.0912 [hep-th]]}.
\bibitem{1406.0505} A. Dabholkar, N. Drukker and J. Gomes, \textit{"Localization in supergravity and quantum $AdS_4/CFT_3$ holography"}, JHEP 1410, 90 (2014), \href{http://arxiv.org/abs/1406.0505}{[arXiv:1406.0505 [hep-th]]}.
\bibitem{1406.1892} D. Sh. Li, Z. W. Liu, J. B. Wu and B. Chen, \textit{"M5-branes in AdS$_4 \times Q^{1,1,1}$ spacetime"}, Phys. Rev. D 90, 066005 (2014), \href{http://arxiv.org/abs/1406.1892}{[arXiv:1406.1892 [hep-th]]}.
\bibitem{1411.6170} E. D'Hoker, \textit{"Exact M-theory solutions, integrable systems, and superalgebras"}, SIGMA 11, 029 (2015), \href{http://arxiv.org/abs/1411.6170}{[arXiv:1411.6170 [hep-th]]}.
\bibitem{1407.3786} Y. Hatsuda and K. Okuyama, \textit{"Probing non-perturbative effects in M-theory"}, JHEP 1410, 158 (2014), \href{http://arxiv.org/abs/1407.3786}{[arXiv:1407.3786 [hep-th]]}.
\bibitem{1409.1799} S. Codesido, A. Grassi and M. Marino, \textit{"Exact results in $\mathcal{N}$=8 Chern-Simons-matter theories and quantum geometry"}, JHEP 1507, 011 (2015), \href{http://arxiv.org/abs/1409.1799}{[arXiv:1409.1799 [hep-th]]}.
\bibitem{1507.00442} H. Ouyang, J. B. Wu and J. J. Zhang, \textit{"Exact results for Wilson loops in orbifold ABJM theory"}, \href{http://arxiv.org/abs/1507.00442}{[arXiv:1507.00442 [hep-th]]}.
\bibitem{1411.3335} Y. Bea, N. Jokela, M. Lippert, A. V. Ramallo and D. Zoakos, \textit{"Flux and Hall states in ABJM with dynamical flavors"}, JHEP 1503, 009 (2015), \href{http://arxiv.org/abs/1411.3335}{[arXiv:1411.3335 [hep-th]]}.

\end{thebibliography}
\end{document}